\documentclass[12pt,english]{article}
\usepackage[T1]{fontenc}
\usepackage{amsmath}
\usepackage{amssymb}

\makeatletter

\newcommand{\noun}[1]{\textsc{#1}}

 \usepackage{amsthm, amsmath, amsfonts, amssymb}
 \theoremstyle{plain}    
 \newtheorem{thm}{Theorem}

 \theoremstyle{remark}    
 \newtheorem*{acknowledgement*}{Acknowledgement} 
 \theoremstyle{plain}    
 \newtheorem{cor}[thm]{Corollary} 
 \theoremstyle{plain}    
 \newtheorem{prop}[thm]{Proposition} 

 \theoremstyle{plain}    
 \newtheorem{lem}{Lemma} 

\usepackage{a4wide}

\def\defi{\stackrel{\rm def}{=}}

\usepackage{color}
\usepackage{graphics}

\usepackage{babel}
\makeatother
\begin{document}

\title{Ruelle-Pollicott resonances for real analytic hyperbolic maps}

\author{Frédéric Faure\textit{}%
\thanks{Institut Fourier 100, rue des Maths, BP 74 ,38402 St Martin d'Heres,
France. \protect \\
email: frederic.faure@ujf-grenoble.fr \protect \\
http://www-fourier.ujf-grenoble.fr/\textasciitilde{}faure%
}\textit{}\\
Nicolas Roy%
\thanks{Geometric Analysis group, Institut für Mathematik, Humbold Universität,
Berlin. roy@math.hu-berlin.de%
}}

\maketitle
\begin{abstract}
We study two simple real analytic uniformly hyperbolic dynamical systems:
expanding maps on the circle $S^{1}$ and hyperbolic maps on the torus
$\mathbb{T}^{2}$. We show that the Ruelle-Pollicott resonances which
describe time correlation functions of the chaotic dynamics can be
obtained as the eigenvalues of a trace class operator in Hilbert space
$L^{2}\left(S^{1}\right)$ or $L^{2}\left(\mathbb{T}^{2}\right)$
respectively. The trace class operator is obtained by conjugation
of the Ruelle transfer operator in a similar way quantum resonances
are obtained in open quantum systems. We comment this analogy.

\end{abstract}
\tableofcontents{}

\paragraph{}

\section{Introduction}

In this paper, we study the Ruelle-Pollicott resonances of two simple
models which are uniformly hyperbolic: expanding maps on the circle
$S^{1}$ and hyperbolic maps on the torus $\mathbb{T}^{2}$. These
models are the most simple examples of chaotic dynamical systems,
which exhibit strong chaotic properties, such as ergodicity, mixing,
decay of correlations, central limit theorem for observables, etc.
, see \cite{katok_hasselblatt}\cite{gaspard-98}\cite{baladi_livre_00}.
Expansivity or hyperbolicity makes every trajectory unstable and are
therefore important hypothesis responsible for these chaotic properties.
One of these properties, the {}``decay of time correlation'' is
fundamental to establish other chaotic properties. Decay of time correlations
can be studied by means of the spectral analysis of {}``Ruelle transfer
operators'', which transport functions on $S^{1}$ or $\mathbb{T}^{2}$
according to the dynamics \cite{eckmann_ruelle_85,ruelle_86}. The
simplest Ruelle transfer operator is the {}``pull back operator''
or {}``Koopman operator'' defined by: $\left(\hat{M}\varphi\right)\left(x\right)\defi\varphi\left(M\left(x\right)\right)$,
where $M:S^{1}\rightarrow S^{1}$ is the map (resp. $M:\mathbb{T}^{2}\rightarrow\mathbb{T}^{2}$\noun{)},
and $\varphi\in C^{0}\left(S^{1}\right)$\noun{.} The time correlation
of two functions $\phi,\varphi\in C^{0}\left(S^{1}\right)$ is defined
by $C_{\phi,\varphi}\left(t\right)\defi\langle\phi|\hat{M}^{t}|\varphi\rangle=\int_{S^{1}}\phi\left(x\right)\left(\varphi\circ M^{t}\left(x\right)\right)dx$.
The main effect of the hyperbolic dynamics is to transform an initial
function $\varphi$ into a function $\hat{M}^{t}\varphi$ with finer
and finer structures, as the time $t$ evolves. In other words, the
information on the initial function $\varphi$ is sent towards {}``microscopic
scales'', or equivalently at infinity in the Fourier space. On the
{}``macroscopic scale'' (i.e., if $\hat{M}^{t}\varphi$ is tested
on a regular function $\phi$), there remains only a constant function,
i.e., the function $1$ times a weight $\mu_{SRB}\left(\varphi\right)$
where $\mu_{SRB}$ is called the Sinai-Ruelle-Bowen (S.R.B.) measure.
The number $\mu_{SRB}\left(\varphi\right)\in\mathbb{C}$ is the only
information on $\varphi$ which remains on the macroscopic scale after
a long time evolution. With this point of view, the decay of time
correlation functions $C_{\phi,\varphi}\left(t\right)$ is due to
the escape of the function $\hat{M}^{t}\varphi$ towards infinity
in the Fourier space, implying a decay of the small Fourier components.
This suggests to study the decay of correlations using a {}``window
of observation'' in the Fourier space, centered on small Fourier
components (in the unstable direction). This is the role of the operator
$\hat{A}$ below. This situation is very similar to open quantum systems,
where the decay of the quantum wave function in a compact domain of
space is due to the escape of the wave function towards infinity.
In such situations people study the decay by a {}``complex scaling
method'' which consists in conjugating the dynamical operator in
such a way that the wave function is toned down at infinity (\cite{simon_87},
chap. 8). Then, the {}``quantum resonances'' which appear are suitable
to describe the decay. Here we will follow this general approach.

We will suppose in both models that the map $M$ is \emph{real analytic}.
We will show that the time correlation functions can be obtained from
an {}``effective dynamical operator'' $\hat{R}$ obtained from $\hat{M}$
by a conjugation $\hat{R}=\hat{A}\hat{M}\hat{A}^{-1}$, where $\hat{A}$
tones down the high Fourier modes in the unstable direction of the
dynamics. The main result is to show that $\hat{R}$ is a trace class
operator in the Hilbert space $L^{2}\left(S^{1}\right)$ (resp. $L^{2}\left(\mathbb{T}^{2}\right)$).
The effective long time dynamics is obtained by $\hat{R}^{t}=\hat{A}\hat{M}^{t}\hat{A}^{-1}$
with $t\in\mathbb{N}$, and the spectral properties of $\hat{R}$
are important for that. The eigenvalues of $\hat{R}$ are called the
{}``Ruelle-Pollicott resonances''. This approach, with a conjugation,
is the one which is usually followed in the complex scaling methods
\cite{simon_87}. An alternative but equivalent approach would be
to keep the operator $\hat{M}$ but to work with another norm instead
of the $L^{2}$ one, namely with $\left(\phi,\psi\right)_{A}\defi\left(\hat{A}\phi,\hat{A}\psi\right)_{L^{2}}$.
This last formulation is preferred in \cite{liverani_02}\cite{liverani_04}\cite{liverani_05}\cite{baladi_05}.
In this approach, the operator $\hat{A}$ is seen as an isomorphism
between two Hilbert spaces, and one gets $\left(\phi,\hat{M}\psi\right)_{A}=\left(\phi',\hat{R}\psi'\right)_{L^{2}}$
with $\phi'=\hat{A}\phi$, $\psi'=\hat{A}\psi$. 

The correlation spectrum for analytic maps has already been studied
through trace class operators but with different approaches: the case
of expanding maps has been studied by D. Ruelle in \cite{ruelle_86}.
The case of analytic hyperbolic maps has been studied by H.H. Rugh
in \cite{rugh_92}. Our approach which consists in working in the
Fourier space (or more technically conjugating $\hat{M}$ by a pseudo-differential
operator) has been already investigated recently for hyperbolic diffeomorphisms
by V. Baladi and M. Tsujii \cite{baladi_05}. In their paper, they
consider a much broader class of dynamical systems (they do not suppose
analyticity) and they show quasi-compacity for the transfer operator
in a Banach space of distributions. Although our results are more
restrictive, we believe that they have their own interest because
of their technical simplicity (we obtain a trace class operator in
a Hilbert space and the proof is quite simple). A technical difference
between the two approaches appears for example with the choice of
the operator $\hat{A}$. We have to choose an operator which has an
exponential expression in the Fourier basis, whereas in \cite{baladi_05}
the operator $\hat{A}$ has an algebraic dependence on the Fourier
modes ($\hat{A}$ is a power of the Laplacian). This difference is
crucial to obtain some of the results presented in this paper. 

Our techniques do not allows us to treat any hyperbolic maps on the
torus, but maps which are closed enough to the linear hyperbolic map%
\footnote{We discuss some possible extension in the conclusion%
}. More precisely, as the proof will show in Section \ref{sub:Proof-of-Theorem_T2},
the technical assumption is that the unstable and stable tangent directions
are uniformly contained in constant cones defined by the linear map.
This restriction of our method, due to the very simple expression
of the operator $\hat{A}$, saves us on the other hand from making
a partition of the unity as in \cite{baladi_05}. 

With a different approach, results close to those obtained by V. Baladi
et al. were obtained by C. Liverani et al. in \cite{liverani_02}\cite{liverani_04}\cite{liverani_05}.
The connection between dynamical determinants and Ruelle resonances
is established there in great generality. See \cite{baladi_05} for
historical remarks and comparison between the different approaches. 

The paper is organized as follows. In Section 2, we define the operator
$\hat{R}$ which governs the decay of the correlation function. We
state the theorems which say that $\hat{R}$ is a trace class operator
in both case $S^{1}$ and $\mathbb{T}^{2}$. These results are proved
respectively in Section 3 and 4. Our method is very convenient for
numerical analysis since the truncation of the high Fourier modes
produces a small error. In Section 3.3, we present numerical illustrations
of some aspects of the S.R.B. measure and the R.P. resonances. In
Section \ref{sub:Relation-with-randomly}, we show the equivalence
of our approach with a more common approach known as {}``randomly
perturbed dynamics'' or {}``noisy models'' \cite{baladi_03}\cite{liverani_02}.
We use this equivalence to show the (well known) Trace formula in
terms of periodic orbits \cite{baladi_livre_00}. We mention that
in \cite{liverani_04}, a powerful method is developed in the $C^{r}$
case and allows the authors to show spectral stability for a wider
class of deterministic and random perturbations.

\begin{acknowledgement*}
We would like to thank Viviane Baladi, Patrick Bernard, Alain Joye,
Malik Mezzadri and Stéphane Nonnenmacher for discussions related to
this work. FF gratefully acknowledges partial support by {}``Agence
Nationale de la Recherche'' under the grant JC05\_52556.
\end{acknowledgement*}

\section{Statement of the results}

\subsection{Expanding map on the circle}

Let $M:\mathbb{R}\rightarrow\mathbb{R}$ be the map defined by%
\footnote{More generally we could have supposed that $M\left(x\right)=dx+f\left(x\right)$
with $d\in\mathbb{N}$, $d\geq2$. This does not change the results
we obtain.%
}\[
M\left(x\right)=2x+f\left(x\right)\]
where $f$ is \emph{real analytic} and periodic: $f\left(x+1\right)=f\left(x\right),\forall x\in\mathbb{R}$.
We suppose moreover that \[
f'_{\textrm{min}}\defi\min_{x\in\mathbb{R}}\left(\frac{df}{dx}\right)>-1\]
 so that $M'_{x}=2+\frac{df}{dx}>1$ ($M$ is called strictly expanding).
A simple example used later for numerical illustrations is \begin{equation}
f\left(x\right)=\frac{\delta}{2\pi}\sin\left(2\pi x\right),\qquad\left|\delta\right|<1\label{eq:example_f}\end{equation}

For all $n\in\mathbb{N}$ and all $x\in\mathbb{R},\, M\left(x+n\right)\equiv M\left(x\right)\,\textrm{mod}1$,
hence $M$ defines an expanding map on the circle $S^{1}=\mathbb{R}/\mathbb{Z}$
(also denoted by $M$). $M$ is not invertible, but it is the simplest
model exhibiting chaotic dynamics, see \cite{katok_hasselblatt}:
expansivity is responsible for the high sensitivity to initial conditions,
mixing and positive entropy of the dynamics.

The pull-back operator $\hat{M}$ on $L^{2}\left(S^{1}\right)$ is
the (non unitary) bounded operator defined by%
\footnote{We can also consider a more general class of operators, called {}``Ruelle
transfer operator'': $\left(\hat{M}_{g}\varphi\right)\left(x\right)\defi e^{g\left(x\right)}\varphi\left(Mx\right)$,
with complex valued function $g$. The results obtained in this paper
extend to this case provided $g$ is real analytic.%
}\begin{equation}
\left(\hat{M}\varphi\right)\left(x\right)\defi\varphi\left(Mx\right)\label{eq:M_on_S1}\end{equation}

For any $n\in\mathbb{Z},$ we denote%
\footnote{Throughout the paper, we use \textbf{Dirac notations} for vectors
in Hilbert space $\mathcal{H}=L^{2}\left(S^{1}\right)$: $\varphi\in\mathcal{H}$
is written $|\varphi\rangle$. Its dual metric is written $\langle\varphi|$.
A scalar product is written $\langle\phi|\varphi\rangle=\int_{S^{1}}\overline{\phi\left(x\right)}\varphi\left(x\right)dx$.
If $\hat{M}$ is an operator, we write $\langle\phi|\hat{M}|\varphi\rangle\defi\langle\phi|\hat{M}\varphi\rangle=\langle\hat{M}^{\ast}\phi|\varphi\rangle$,
(where $\hat{M}^{\ast}$ is the adjoint operator). Finally, $|\phi\rangle\langle\varphi|\defi|\phi\rangle\otimes\langle\varphi|$
. %
} by $|n\rangle\in L^{2}\left(S^{1}\right)$ the Fourier mode $\varphi_{n}\left(x\right)=\exp\left(i2\pi nx\right)$,
and define the operator $\hat{A}$ by\begin{equation}
\hat{A}|n\rangle\defi\exp\left(-a\left|n\right|\right)|n\rangle,\qquad n\in\mathbb{Z},\qquad\textrm{with }a>0.\label{eq:def_A_S1}\end{equation}
$\hat{A}$ is diagonal in the Fourier basis. The image $C_{A}=\hat{A}\left(L^{2}\left(S^{1}\right)\right)$
is a set of very regular functions (analytic in a complex neighborhood
of $S^{1}$ of radius $a$). The operator $\hat{A}$ is used to define%
\footnote{Equivalently $\hat{A}$ can be seen as a change of norm on $L^{2}\left(S^{1}\right)$. %
} the operator:\begin{equation}
\hat{R}\defi\hat{A}\hat{M}\hat{A}^{-1}\label{eq:def_R_S1}\end{equation}
with domain $C_{A}$, dense in $L^{2}\left(S^{1}\right)$. 

In this paper, we will prove the following theorem:

\begin{center}\fbox{\parbox{16cm}{

\begin{thm}
\label{thm:R_S1}There exists $a>0$ entering in eq.(\ref{eq:def_A_S1}),
such that matrix elements of $\hat{R}$ decrease exponentially: $\left|\langle n'|\hat{R}|n\rangle\right|<\exp\left(-c\left(\left|n'\right|+\left|n\right|\right)\right)$,
with $c>0$. In particular $\hat{R}$ extends to a trace class operator
in Hilbert space $L^{2}\left(S^{1}\right)$ .
\end{thm}
}}\end{center}

As the proof will show, the result holds for any $a\in\left]0,a_{0}\right]$,
with some  $a_{0}>0$, and $c$ depends on $a$.

For general results on trace class operators, see \cite{reed-simon1,reed-simon4},
or \cite{gohberg-00} chap. 4. The eigenvalues of $\hat{R}$ are called
the \textbf{Ruelle-Pollicott resonances} (R.P.) of the pull back operator
$\hat{M}$.

\subsubsection{Dynamical correlation functions}

The operator $\hat{R}$ is useful to study time-correlation functions.
Indeed, if $|\phi\rangle\in C_{A}$ is a regular {}``test function'',
and $|\varphi\rangle\in L^{2}\left(S^{1}\right)$, then for $t\in\mathbb{N}$,
$C_{\phi,\varphi}\left(t\right)\defi\langle\phi|\hat{M}^{t}|\varphi\rangle$
can be expressed using $\hat{R}$ as\begin{equation}
C_{\phi,\varphi}\left(t\right)\defi\langle\phi|\hat{M}^{t}|\varphi\rangle=\left(\langle\phi|\hat{A}^{-1}\right)\hat{R}^{t}\left(\hat{A}|\varphi\rangle\right)\label{eq:Corr_S1}\end{equation}
In {}``physical terms'' it means that the operator $\hat{R}$ is
a nice effective operator to express the dynamics of $\hat{M}$ in
$L^{2}\left(S^{1}\right)$ provided it is tested on the regular function
space $C_{A}$.

\subsubsection{Finite rank approximation}

There is a direct consequence of Theorem \ref{thm:R_S1} which is
useful for the numerical computation of Ruelle-Pollicott resonances.
Let $\hat{M}_{N}$ be the matrix of the operator $\hat{M}$ expressed
in the Fourier basis and truncated to the $N$ first Fourier modes
(i.e., $\left|n\right|\leq N$ , the matrix $\hat{M}_{N}$ has thus
size $\left(2N+1\right)\times\left(2N+1\right)$).

\begin{cor}
The eigenvalues of $\hat{M}_{N}$ converge towards the Ruelle-Pollicott
resonances, when $N\rightarrow\infty$.
\end{cor}
\begin{proof}
If $\hat{R}_{N}$ is the matrix of the operator $\hat{R}$ restricted
to $\mathcal{H}_{N}=\textrm{Span}\left\{ |n\rangle,\quad\left|n\right|\leq N\right\} \equiv\mathbb{C}^{2N+1}$,
then the eigenvalues of $\hat{R}_{N}$ converge to the R.P. resonances
as $N\rightarrow\infty$, because $\hat{R}_{N}$ converges to $\hat{R}$
in operator norm. But in $\mathcal{H}_{N}\equiv\mathbb{C}^{N+1}$,
$\hat{R}_{N}$ is conjugate to $\hat{M}_{N}$ by the invertible and
diagonal matrix $\hat{A}_{N}=\textrm{Diag}\left(\exp\left(-a\left|n\right|\right),n=-N\rightarrow N\right)$,
$\hat{R}_{N}$ and $\hat{M}_{N}$ have the same spectrum.
\end{proof}

\subsubsection{Exponential concentration of R.P. resonances near zero}

Since $\hat{R}$ is a compact operator (and moreover a trace class
operator), its eigenvalues converge to zero. From Theorem \ref{thm:R_S1},
matrix elements $\langle n'|\hat{R}|n\rangle$ decrease exponentially
fast for large $\left|n\right|,\left|n'\right|$. A quite direct consequence
of this is the exponential concentration of the eigenvalues of $\hat{R}$
near zero:

\begin{thm}
\label{th:exp_accumulation}Let $\lambda_{i}\in\mathbb{C}$, $i=0,1,\ldots$
be the non zero eigenvalues of $\hat{R}$ (the Ruelle-Pollicott resonances),
such that $\left|\lambda_{i+1}\right|\leq\left|\lambda_{i}\right|$,
and counting multiplicity. Let $l_{i}=\log\left|\lambda_{i}\right|$
and $C_{1}=\frac{2\left(1+e^{-c}\right)}{\left(1-e^{-c}\right)^{2}}$.
For any $i\geq0$,\begin{equation}
l_{i}\leq-\frac{c}{4}i+\log C_{1}\label{eq:accumul_li_S1}\end{equation}

\end{thm}
The constant $c$ is given in Theorem \ref{thm:R_S1} and the proof
of this theorem is postponed to Section \ref{sub:Exponential-accumulation-of}.
(The best estimate is for the largest possible value of $c$, which
is neither very explicit in the proof, nor very sharp).

\subsubsection{Relation with randomly perturbed operators or noisy models\label{sub:Relation-with-randomly}}

A different approach to obtain Ruelle-Pollicott resonances of transfer
operators is to add a small {}``noise'', or {}``random perturbation''
to the dynamical operator $\hat{M}$ at each step of evolution. These
models are often used for theoretical or numerical calculations. In
\cite{baladi_03} Baladi and Young show that for expanding maps, the
randomly perturbed operator is compact, and that its eigenvalues are
the Ruelle-Pollicott resonances in the limit when the perturbation
vanishes. A similar result is shown by Blank Keller and Liverani in
\cite{liverani_02} for Anosov maps. See also \cite{nonnenmacher-noisy-03}.
In this section we consider such {}``noisy operators'' and show
with a very simple proof that they have the same eigenvalues as $\hat{R}$
(i.e., the R.P. resonances) when the perturbation vanishes. The same
result holds for Anosov maps on $\mathbb{T}^{2}$ considered in the
next section.

Let $\Delta\equiv-\frac{d^{2}}{dx^{2}}$ be the Laplacian operator
on $S^{1}$, and for $\varepsilon>0$, let \[
\hat{\mathcal{D}}_{\varepsilon}\defi e^{-\varepsilon\Delta/\left(2\pi\right)^{2}}\]
be the heat operator. This operator is diagonal in the Fourier basis:
\begin{equation}
\hat{\mathcal{D}}_{\varepsilon}|n\rangle=e^{-\varepsilon n^{2}}|n\rangle\label{eq:def_De_S1}\end{equation}

The main effect of $\hat{\mathcal{D}}_{\varepsilon}$ is to truncate
the high Fourier components. In the {}``real space'' $x\in S^{1}$,
the operator $\hat{\mathcal{D}}_{\varepsilon}$ convolutes with a
Gaussian distribution of size $\sim\sqrt{\varepsilon}$, so $\hat{\mathcal{D}}_{\varepsilon}$
as the same effect as a Gaussian noise.

Let us define the {}``noisy perturbed operator'' by\begin{equation}
\hat{M}_{\varepsilon}\defi\hat{M}\hat{\mathcal{D}}_{\varepsilon}\label{eq:def_M_eps}\end{equation}
 $\hat{M}_{\varepsilon}$ is a Trace class operator because it is
the product of $\hat{\mathcal{D}}_{\varepsilon}$ which is Trace class
with $\hat{M}$ which is bounded (cf \cite{reed-simon1} p.207).

\begin{thm}
\label{thm:spectre_de_M_eps}The eigenvalues of the noisy perturbed
operator $\hat{M}_{\varepsilon}$ converge to the Ruelle-Pollicott
resonances, when $\varepsilon\rightarrow0$.
\end{thm}
\begin{proof}
Let us define the operator\[
\hat{R}_{\varepsilon}\defi\hat{R}\hat{\mathcal{D}}_{\varepsilon}=\hat{A}\hat{M}\hat{A}^{-1}\hat{\mathcal{D}}_{\varepsilon}=\hat{A}\hat{M}\hat{\mathcal{D}}_{\varepsilon}\hat{A}^{-1}=\hat{A}\hat{M}_{\varepsilon}\hat{A}^{-1}\]
where we have used the fact that $\hat{A}$ and $\hat{\mathcal{D}}_{\varepsilon}$
commute. Let $\hat{R}_{\varepsilon,N}$ (resp. $\hat{M}_{\varepsilon,N}$)
the matrix of the operator $\hat{R}_{\varepsilon}$ (resp. $\hat{M}_{\varepsilon}$)
expressed in the Fourier basis and truncated to the first $N$ Fourier
modes. We have $\hat{R}_{\varepsilon,N}=\hat{A}\hat{M}_{\varepsilon,N}\hat{A}^{-1}$
so the matrices $\hat{R}_{\varepsilon,N}$ and $\hat{M}_{\varepsilon,N}$
have the same spectrum. But $\hat{R}_{\varepsilon,N}$ converges to
$\hat{R}_{\varepsilon}$ in operator norm for $N\rightarrow\infty$
(resp. $\hat{M}_{\varepsilon,N}\rightarrow\hat{M}_{\varepsilon}$
for $N\rightarrow\infty$). We deduce that $\hat{R}_{\varepsilon}$
and $\hat{M}_{\varepsilon}$ have the same spectrum. Now $\hat{R}_{\varepsilon}$
converges to $\hat{R}$ in operator norm for $\varepsilon\rightarrow0$,
which gives that eigenvalues of $\hat{R}_{\varepsilon}$ converge
to eigenvalues of $\hat{R}$.
\end{proof}

\subsubsection{Trace formula}

An important feature of Ruelle transfert operators is the existence
of exact trace formulas in terms of periodic orbits (\cite{baladi_livre_00}
p. 100). We just recall here this trace formula for the operator $\hat{R}$
defined by Eq.(\ref{eq:def_R_S1}).

\begin{prop}
\label{prop:Trace_formula_S1}For any $t\geq1$,\[
\textrm{Tr}\left(\hat{R}^{t}\right)=\sum_{x\in\textrm{Fix}\left(M^{t}\right)}\frac{1}{\left|D_{x}M^{t}-1\right|}\]
where $\textrm{Fix}\left(M^{t}\right)$ denotes the set of fixed points
of $M^{t}$, and $D_{x}M^{t}\left(x\right)=\frac{dM^{t}}{dx}\left(x\right)$.
\end{prop}
\begin{proof}
We consider first the trace of the operator $\hat{M}_{t,\varepsilon}\defi\hat{M}^{t}\hat{D}_{\varepsilon}$,
(similar to Eq.(\ref{eq:def_M_eps})). From (\cite{gohberg-00} Th.
8.1 p. 70), \[
\textrm{Tr}\left(\hat{M}_{t,\varepsilon}\right)=\int_{0}^{1}dx\langle x|\hat{M}_{t,\varepsilon}|x\rangle=\int_{0}^{1}dx\delta_{\varepsilon}\left(M^{t}\left(x\right)-x\right)\]
where $\langle x'|\hat{M}_{t,\varepsilon}|x\rangle$ denotes the Schwartz
kernel of the operator $\hat{M}_{t,\varepsilon}$ on $L^{2}\left(S^{1}\right)$
and where $\delta_{\varepsilon}=\hat{D}_{\varepsilon}\delta$ denotes
the regularized Dirac distribution at $x=0$ (i.e. $\delta_{\varepsilon}$
is a periodic Gaussian function with width $\sim\sqrt{\varepsilon}$).
With the ($\left(2^{t}-1\right)$-valued) change of variable $y=M^{t}\left(x\right)-x$,
we obtain\[
\textrm{Tr}\left(\hat{M}_{t,\varepsilon}\right)=\int_{0}^{2^{t}-1}dy\frac{1}{\left|D_{x}M^{t}-1\right|}\delta_{\varepsilon}\left(y\right)\]
so $\textrm{Tr}\left(\hat{M}_{t,\varepsilon}\right)\rightarrow\sum_{x\in\textrm{Fix}\left(M^{t}\right)}\frac{1}{\left|D_{x}M^{t}-1\right|}$
for $\varepsilon\rightarrow0$. On the other hand, with $\hat{R}_{t,\varepsilon}\defi\hat{R}^{t}\hat{\mathcal{D}}_{\varepsilon}$,
we show (as in the proof of Theorem \ref{thm:spectre_de_M_eps}),
that $\textrm{Tr}\left(\hat{M}_{t,\varepsilon}\right)=\textrm{Tr}\left(\hat{R}_{t,\varepsilon}\right)$,
and that $\textrm{Tr}\left(\hat{R}_{t,\varepsilon}-\hat{R}^{t}\right)\rightarrow0$,
for $\varepsilon\rightarrow0$. This gives the result.
\end{proof}

\subsection{Hyperbolic map on the torus}

With the same approach, one can study a non linear hyperbolic map
on the torus (expressed as a linear map with a small perturbation).
Let $M_{0}\in SL\left(2,\mathbb{Z}\right)$ be an hyperbolic matrix
(i.e., $\textrm{Tr}M_{0}>2$) and $f:\mathbb{R}^{2}\rightarrow\mathbb{R}^{2}$
be a real analytic periodic function:\[
f\left(x+n\right)=f\left(x\right),\forall x\in\mathbb{R}^{2},\forall n\in\mathbb{Z}^{2}\]
Let $M:\mathbb{R}^{2}\rightarrow\mathbb{R}^{2}$ be defined to be
$M_{0}$ perturbed by $f$:\[
M\left(x\right)=M_{0}\left(x\right)+f\left(x\right)\]
Then $M\left(x+n\right)\equiv M\left(x\right)\,\left[\mathbb{Z}^{2}\right]$
hence $M$ induces a map on $\mathbb{T}^{2}$ also denoted by $M$.
Structural stability asserts that the map $M$ on $\mathbb{T}^{2}$
is Anosov (uniformly hyperbolic) whenever $\left\Vert f\right\Vert _{C^{1}}$
is small enough (\cite{arnold-ed2} p.122)(\cite{katok_hasselblatt}
p. 89)%
\footnote{In our case $f$ is supposed to be real analytic. So $\left\Vert f\right\Vert _{C^{1}}$
is controlled by $\left\Vert f\right\Vert _{C^{0}}$.%
}.

The pull-back operator $\hat{M}$ on $L^{2}\left(\mathbb{T}^{2}\right)$
is the bounded operator defined by\begin{equation}
\left(\hat{M}\varphi\right)\left(x\right)\defi\varphi\left(Mx\right).\label{eq:M_hat_T2}\end{equation}
Note that the operator $\hat{M}$ is not unitary except if $M$ preserves
the area on $\mathbb{T}^{2}$.

\paragraph{Example:}

Choose $M_{0}=\left(\begin{array}{cc}
2 & 1\\
1 & 1\end{array}\right)$ and \begin{equation}
f\left(x\right)=\left(0,\frac{\delta}{2\pi}\sin\left(2\pi\left(2x_{1}+x_{2}\right)\right)\right).\label{eq:example_f_T2}\end{equation}
In this example, $M$ preserves%
\footnote{Because in this example, $M$ can be written as $M=M_{1}M_{0}$, where
$M_{1}$ is an Hamiltonian time 1 flow, generated by Hamiltonian function
$H_{1}\left(x_{1},x_{2}\right)=\frac{\delta}{\left(2\pi\right)^{2}}\cos\left(2\pi x_{1}\right)$
on $\mathbb{T}^{2}$.%
} area $dx_{1}dx_{2}$.

For each $n=\left(n_{1},n_{2}\right)\in\mathbb{Z}^{2}$, denote by
$|n\rangle\in L^{2}\left(\mathbb{T}^{2}\right)$ the Fourier mode
$\varphi_{n}\left(x\right)=\exp\left(i2\pi\left(n.x\right)\right)$.
Let $u,s\in\mathbb{R}^{2}$ be unstable/stable eigenvectors of the
transposed matrix $M_{0}^{\mathfrak{t}}$, i.e., $M_{0}^{\mathfrak{t}}u=e^{\lambda_{0}}u$
and $M_{0}^{\mathfrak{t}}s=e^{-\lambda_{0}}s$, with $\lambda_{0}>0$.
A vector $v=\left(v_{1},v_{2}\right)\in\mathbb{R}^{2}$ is written
$\tilde{v}\equiv\left(v_{u},v_{s}\right)$ with respect to the basis
$\left(u,s\right)$. In particular $n\in\mathbb{Z}^{2}$ is mapped
to $\tilde{n}=\left(n_{u},n_{s}\right)$. Define the operator $\hat{A}$
by\begin{equation}
\hat{A}|n\rangle\defi\exp\left(-a\left|n_{u}\right|+a\left|n_{s}\right|\right)|n\rangle,\qquad n\in\mathbb{Z}^{2},\quad\textrm{with }a>0.\label{eq:def_A_T2}\end{equation}

It is diagonal in the Fourier basis. $\hat{A}$ is defined on a domain
$D_{A}\defi\textrm{Dom}\left(\hat{A}\right)=\left\{ |\varphi\rangle=\sum_{n}\varphi_{n}|n\rangle,\,\, s.t.\,\,\sum_{n}\left|\varphi_{n}\right|^{2}e^{2a\left|n_{s}\right|-2a\left|n_{u}\right|}<\infty,\,\sum_{n}\left|\varphi_{n}\right|^{2}<\infty\right\} $
dense in $L^{2}\left(\mathbb{T}^{2}\right)$, and consists of functions
with exponentially fast decreasing Fourier components (i.e very regular)
in the stable direction. Similarly,\begin{eqnarray*}
C_{A}\defi\textrm{Dom}\left(\hat{A}^{-1}\right) & = & \left\{ |\phi\rangle=\sum_{n}\phi_{n}|n\rangle,\,\, s.t.\,\,\sum_{n}\left|\phi_{n}\right|^{2}e^{-2a\left|n_{s}\right|+2a\left|n_{u}\right|}<\infty,\,\sum_{n}\left|\phi_{n}\right|^{2}<\infty\right\} \\
 &  & \subset L^{2}\left(\mathbb{T}^{2}\right)\end{eqnarray*}
 consists of functions with exponentially fast decreasing Fourier
components in the unstable direction. One checks that $C_{A}=\hat{A}\left(D_{A}\right)$,
$D_{A}=\hat{A}^{-1}\left(C_{A}\right)$.

Define\begin{equation}
\hat{R}\defi\hat{A}\hat{M}\hat{A}^{-1}\label{eq:def_R_T2}\end{equation}
on a suitable domain included in $C_{A}$ (one has $\textrm{Dom}\left(\hat{R}\right)=C_{A}$
if $\hat{M}\left(D_{A}\right)\subset D_{A}$).

\begin{center}\fbox{\parbox{16cm}{

\begin{thm}
\label{thm:M_nucleaire_T2}For a $C^{1}$-small enough perturbation
$f$, i.e., $\left\Vert f\right\Vert _{C^{1}}<\varepsilon$ with $\varepsilon>0$,
there exists $a>0$ such that $\hat{R}$ is defined on the domain
$\textrm{Dom}\left(\hat{R}\right)=C_{A}$ and its matrix elements
decrease exponentially: $\left|\langle n'|\hat{R}|n\rangle\right|<\exp\left(-c\left(\left|n_{1}'\right|+\left|n_{2}'\right|+\left|n_{1}\right|+\left|n_{2}\right|\right)\right)$,
with $c>0$. Therefore, $\hat{R}$ extends to a trace class operator
in Hilbert space $L^{2}\left(\mathbb{T}^{2}\right)$ .
\end{thm}
}}\end{center}

As the proof will show, the result holds for any $a\in\left]0,a_{0}\right]$,
with some $a_{0}>0$, and $c$ depends on $a$. The operator $\hat{R}$
is useful to study time-correlation functions: if $|\phi\rangle\in C_{A}$
is a regular {}``test function'', and $|\varphi\rangle\in D_{A}$,
then\[
C_{\phi,\varphi}\left(t\right)\defi\langle\phi|\hat{M}^{t}|\varphi\rangle=\left(\langle\phi|\hat{A}^{-1}\right)\hat{R}^{t}\left(\hat{A}|\varphi\rangle\right)\]

The different corollaries and applications we mentionned for expanding
maps, work as well for hyperbolic maps on the torus (with only minor
modifications): (1) finite rank approximation, (2) exponential concentration
of R.P. resonances near zero, (3) relation with randomly perturbed
operators, (4) Trace formula in terms of periodic orbits.

\subsubsection{Exponential concentration of R.P. resonances near zero}

\begin{thm}
\label{th:exp_accumulation_T2}Let $\lambda_{n}\in\mathbb{C}$, $n=0,1,\ldots$
be the non zero eigenvalues of $\hat{R}$ (the Ruelle-Pollicott resonances),
such that $\left|\lambda_{n+1}\right|\leq\left|\lambda_{n}\right|$,
and counting multiplicity. Let $l_{n}=\log\left|\lambda_{n}\right|$.
There exists $C_{1}>0$ such that for any $n\geq0$,\begin{equation}
l_{n}\leq-\frac{c}{3}\sqrt{n}\frac{1}{\left(1+1/n\right)}+\log C_{1}\label{eq:asympt_li_T2}\end{equation}

\end{thm}
the constant $c$ is given in Theorem \ref{thm:R_S1}, and the proof
of the theorem is postponed to Section \ref{sub:Exponential-accumulation-T2}.
The main difference with (\ref{eq:accumul_li_S1}) is the appearance
of $n^{1/2}$, where the power is $1/d$ with $d=\dim\left(\mathbb{T}^{2}\right)=2$.
Let us remark that such an asymptotic behaviour of eigenvalues is
also met in quantum mechanics in the spectrum $\left(E_{n}\right)_{n}$
of the Harmonic Oscillator on $\mathbb{R}^{d}$. From the semi-classical
Weyl law $E_{n}\simeq\mbox{Cste}\, n^{1/d}$ (see \cite{gutzwiller}
chap. 16). This suggests that (\ref{eq:asympt_li_T2}) could be obtained
or interpreted within a semi-classical approach, with a Weyl asymptotic.

\subsubsection{Relation with randomly perturbed operators}

The analysis made in section \ref{sub:Relation-with-randomly}, can
be repeated with no change except for the definition of the heat operator:\[
\hat{\mathcal{D}}_{\varepsilon}|n\rangle=e^{-\varepsilon\left(n_{1}^{2}+n_{2}^{2}\right)}|n\rangle\]
 which is used to define the randomly perturbed operator:

\[
\hat{M}_{\varepsilon}\defi\hat{M}\hat{\mathcal{D}}_{\varepsilon}\]
We have 

\begin{thm}
The eigenvalues of the noisy perturbed operator $\hat{M}_{\varepsilon}$
converge towards the Ruelle-Pollicott resonances, as $\varepsilon\rightarrow0$.
\end{thm}
(The same proof as in section \ref{sub:Relation-with-randomly}.)

\subsubsection{Trace formula}

We have the following Trace formula for $\hat{R}^{t}$ in terms of
periodic orbits. 

\begin{prop}
For any $t\geq1$,\[
\textrm{Tr}\left(\hat{R}^{t}\right)=\sum_{x\in\textrm{Fix}\left(M^{t}\right)}\frac{1}{\left|\textrm{det}\left(D_{x}M^{t}-Id\right)\right|}\]
where $\textrm{Fix}\left(M^{t}\right)$ denotes the set of fixed points
of $M^{t}$, and $D_{x}M^{t}\left(x\right)$ is the differential at
point $x\in\mathbb{T}^{2}$.
\end{prop}
The proof follows the same lines as for Proposition \ref{prop:Trace_formula_S1}.

\section{Expanding map on the circle}

In this section, we prove Theorem \ref{thm:R_S1}.

\subsection{Matrix elements of the operator $\hat{M}$}

Denote $|n\rangle$ the Fourier mode $\varphi_{n}\left(x\right)=\exp\left(i2\pi nx\right)$,
with $n\in\mathbb{Z}$, $x\in S^{1}$. The set $\left(|n\rangle\right)_{n\in\mathbb{Z}}$
form an orthonormal basis of $L^{2}\left(S^{1}\right)$ and matrix
elements of $\hat{M}$ in this basis are explicitly given by\begin{equation}
\langle n'|\hat{M}|n\rangle=\int_{0}^{1}dx\,\exp\left(i2\pi\left(\left(2n-n'\right)x+nf\left(x\right)\right)\right)\label{eq:element_matrice_M_S1}\end{equation}

\subsubsection{Remarks}

\begin{itemize}
\item For a vanishing perturbation $f=0$, then \begin{equation}
\langle n'|\hat{M}_{0}|n\rangle=\delta_{2n=n'}\label{eq:elements_matrice_M0_S1}\end{equation}
 i.e., in the plane $\left(n',n\right)$, matrix elements are zero
except on the {}``line'' $n'=2n$. For a non zero perturbation $f$,
we will show that matrix elements are very small outside a cone containing
this line. 
\item Since $f$ is real, we have the symmetry\[
\langle-n'|\hat{M}|-n\rangle=\overline{\langle n'|\hat{M}|n\rangle}\]
and if $n=0$, we have\[
\langle n'|\hat{M}|0\rangle=\delta_{n'=0}\]
Therefore we have only to study matrix elements with $n>0$.
\end{itemize}

\subsubsection{Localization property of matrix elements}

For simplicity of the presentation, we borrow notations from semi-classical
analysis (see \cite{martinez-01}). For $n>0$, let us make the change
of variables $\left(n,n'\right)\Leftrightarrow\left(h,p\right)$,
with\begin{equation}
h=\frac{1}{n},\qquad p=\frac{1}{n}\left(n'-2n\right)=n'h-2\label{eq:def_h_p}\end{equation}
and define $\hbar=h/\left(2\pi\right)$. A matrix element can be written
as the oscillating integral:\[
I_{\hbar}\left(p\right)\defi\langle n'|\hat{M}|n\rangle=\int_{0}^{1}dx\,\exp\left(i\left(f\left(x\right)-px\right)/\hbar\right)\]

From the {}``non stationary phase theorem'' below, this integral
is {}``very small'' for values of $p$ outside the interval $\left[f'_{\textrm{min}},f'_{\textrm{max}}\right]$,
with \[
f'_{\textrm{min}}\defi\min_{x}\frac{df}{dx},\qquad f'_{\textrm{max}}\defi\max_{x}\frac{df}{dx}.\]

\begin{thm}
\label{thm:Non-stationary-phase.}\textbf{{}``Non stationary phase''}.
Assume that $f\left(x\right)$ is a periodic function and can be continued
analytically in some strip $\left|\textrm{Im}\left(x\right)\right|<Y$.
Assume $p/\left(2\pi\hbar\right)\in\mathbb{Z}$, and $I_{\hbar}\left(p\right)=\int_{0}^{1}dx\exp\left(i\left(f\left(x\right)-px\right)/\hbar\right)$.
For any $\varepsilon>0$, there exists a constant $C>0$, such that
for any $p,\hbar>0$,\begin{equation}
\left|I_{\hbar}\left(p\right)\right|\leq\min\left(1,e^{-C\left(f'_{\textrm{min}}-p-\varepsilon\right)/\hbar},e^{-C\left(p-f'_{\textrm{max}}-\varepsilon\right)/\hbar}\right)\label{eq:majoration_I}\end{equation}

\end{thm}
\begin{figure}[htbp]
\begin{centering}\input{majoration_Ip.pstex_t}\par\end{centering}

\caption{Upper bounds for the function $\left|I_{\hbar}\left(p\right)\right|$}
\end{figure}

\begin{proof}
Write $z=x+iy$ and $f\left(z\right)=a\left(z\right)+ib\left(z\right)$,
with $a,b$ real valued functions. Analyticity of $f$ gives $\partial_{y}b=\partial_{x}a$.
For $y=0$ and $x\in\left[0,1\right]$, one has $b\left(x,0\right)=0$
and $\left(\partial_{y}b\right)\left(x,0\right)=\left(\partial_{x}a\right)_{y=0}\geq f'_{\textrm{min}}$.
Therefore $\forall\varepsilon>0,\exists y_{0}>0$ $y_{0}<Y$ such
that $b\left(x,y_{0}\right)>\left(f'_{\textrm{min}}-\varepsilon\right)y_{0}$
for all $x$. So for $z=x+iy_{0}$, \[
\left|\exp\left(i\left(f\left(z\right)-pz\right)/\hbar\right)\right|=\exp\left(-\left(b\left(x,y_{0}\right)-py_{0}\right)/\hbar\right)<\exp\left(-\left(f'_{\textrm{min}}-\varepsilon-p\right)y_{0}/\hbar\right).\]
Since $f$ is analytic, we can deform the integration path to $z=x+iy_{0}$,
$x=0\rightarrow1$, $y_{0}$ fixed. This gives \[
\left|I_{\hbar}\left(p\right)\right|<\exp\left(-\left(f'_{\textrm{min}}-\varepsilon-p\right)y_{0}/\hbar\right).\]
A similar argument for the integral $\overline{I}_{\hbar}\left(p\right)=\int_{0}^{1}dx\exp\left(i\left(-f\left(x\right)+px\right)/\hbar\right)$
provides\[
\left|I_{\hbar}\left(p\right)\right|<\exp\left(-\left(p-f'_{\textrm{max}}-\varepsilon\right)y_{0}'/\hbar\right)\]
We finally choose $C=\min\left(y_{0},y_{0}'\right)$.
\end{proof}

\subsubsection{Remarks}

\begin{itemize}
\item For $p\in\left[f'_{\textrm{min}},f'_{\textrm{max}}\right]$, the stationary
phase formula gives the asymptotic value of $I_{\hbar}\left(p\right)$
when $\hbar\rightarrow0$. It says that the asymptotic value of the
matrix element $\langle n'|\hat{M}|n\rangle\equiv I_{\hbar}\left(p\right)$
of operator $\hat{M}$ depends only on the point $x'$ in the integral,
such that $p=\frac{df}{dx}\left(x'\right)$, equivalently $n'=\left(2+\frac{df}{dx}\left(x'\right)\right)n$.
We comment now on a semi-classical interpretation of this result.
The map $M^{-1}$ acting in $S^{1}$ is two valued. Let $\tilde{M}^{-1}$
denotes its lifted action on the cotangent space $T^{*}S^{1}$. If
$x=M(x')=2x'+f\left(x'\right)$, then $\frac{\partial}{\partial x'}=\left(2+\frac{df}{dx}\left(x'\right)\right)\frac{\partial}{\partial x}$.
So, if $\left(x,k\right)$ denote coordinates on $T^{*}S^{1}$, and
$\left(x',k'\right)=\tilde{M}^{-1}\left(x,k\right)$, then $x=M\left(x'\right)$
and $k'=\left(2+\frac{df}{dx}\left(x'\right)\right)k$. So in a sense
which need to be precised, $\hat{M}$ acting in $L^{2}\left(S^{1}\right)$
is a {}``semi-classical quantization'' of the map $\tilde{M}^{-1}$
acting in the symplectic space $T^{*}S^{1}$. This {}``semi-classical
aspect'' of hyperbolic dynamics will be investigated in a futur work.
\item If we come back to variables $n,n'$, this last theorem shows that
matrix elements $\langle n'|\hat{M}|n\rangle$ in the plane $\left(n',n\right)$
are exponentially small outside the cone defined by (for $n>0$):\[
\left(f'_{\textrm{min}}+2-\varepsilon\right)n<n'<\left(f'_{\textrm{max}}+2+\varepsilon\right)n\]
Expansivity hypothesis of $M$ gives $f'_{min}+2>1$, so for $\varepsilon$
small enough this cone does not contain the diagonal $n'=n$. Cf Figure
\ref{fig:cone_S1}.%
\begin{figure}[htbp]
\begin{centering}\input{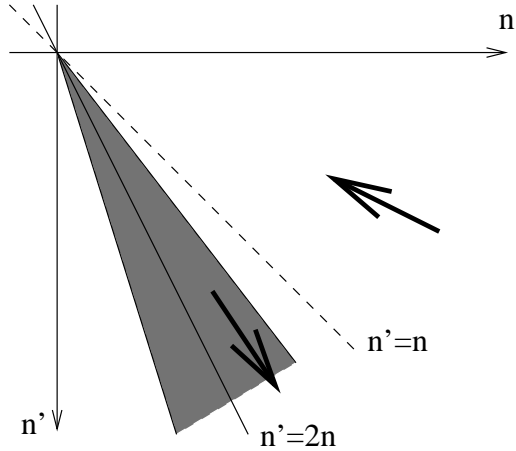}\par\end{centering}

\caption{\label{fig:cone_S1}Matrix elements $\langle n'|\hat{M}|n\rangle$
are small outside the grey cone, for $\left|n\right|\rightarrow\infty$.
For $f=0$, this cone reduces to the line $n'=2n$. These matrix elements
can be interpreted as transition amplitudes for the dynamics $n\rightarrow n'$.
For long times (many iterations), the dynamics goes to infinity in
Fourier space on the sector $\left|n'\right|>\left|n\right|$, and
tends to small Fourier components on the sector $\left|n'\right|<\left|n\right|$
(see the black arrows). Important matrix elements of $\hat{M}$ are
localized on the sector $\left|n'\right|>\left|n\right|$ only, and
thus generate an escape towards infinity in Fourier space, responsible
for chaos, as discussed in the introduction.}
\end{figure}

\item With the example (\ref{eq:example_f}), we can explicitly express
matrix elements in term of Bessel functions of the first kind (see
\cite{abramowitz} 9.1.21 p. 360):\[
\langle n'|\hat{M}|n\rangle=\left(-1\right)^{\left(2n-n'\right)}J_{\left(2n-n'\right)}\left(\delta n\right)\]
This gives\[
\left|I_{\hbar}\left(p\right)\right|=\left|J_{\left(-p/h\right)}\left(\frac{\delta}{h}\right)\right|.\]
Asymptotic results for Bessel functions (see \cite{abramowitz} 9.3.1
and 9.3.2 p. 365) give\begin{eqnarray*}
\log\left|I_{\hbar}\left(p\right)\right| & \sim & -\frac{p}{h}\left(\log\left(\frac{2p}{e\delta}\right)\right),\quad\textrm{for}\, h\,\textrm{fixed, and }p\rightarrow\infty\\
 & \sim & -\frac{p}{h}\left(\alpha-\tanh\alpha\right),\,\textrm{with}\,\cosh\alpha=\frac{p}{\delta}\geq1\textrm{ fixed},\,\textrm{and}\, h\rightarrow0\end{eqnarray*}
which are sharper than the upper bound in (\ref{eq:majoration_I}).
\end{itemize}

\subsection{Proof of Theorem \ref{thm:R_S1}}

\subsubsection{\label{sub:Remarks-313}Idea of the proof and remarks}

In the \emph{linear case}, with a vanishing perturbation $f=0$, the
result is obvious. Indeed from eq.(\ref{eq:elements_matrice_M0_S1}),
matrix elements of $\hat{R}$ lie on the line $n'=2n$ and decrease
like $\langle n'|\hat{R}|n\rangle=\delta_{n'=2n}e^{a\left(\left|n\right|-\left|n'\right|\right)}=\delta_{n'=2n}e^{-a\left|n\right|}$
(the spectrum is then $\sigma\left(\hat{R}\right)=\left\{ 1\right\} \cup\left\{ 0\right\} $,
with $1$ as a simple eigenvalue). Remark that choosing the operator
$\hat{A}$ with the algebraic form $\hat{A}|n\rangle=\frac{1}{\left|n\right|^{\alpha}}|n\rangle,\quad n\in\mathbb{Z}$
would not give decreasing matrix elements: $\langle n'|\hat{R}|n\rangle=\delta_{n'=2n}\frac{\left|n\right|^{a}}{\left|n'\right|^{a}}=\delta_{n'=2n}\frac{1}{2^{a}}$.
On the other hand, the choice $\hat{A}|n\rangle=e^{-\left|n\right|^{a}}|n\rangle$
with $0<a<1$ or $\hat{A}|n\rangle=e^{-a\log^{2}\left(1+\left|n\right|\right)}|n\rangle$
would be suitable as well.

In the \emph{non linear case}, with $f\neq0$, we have shown that
$\left|\langle n'|\hat{M}|n\rangle\right|$ decreases fast outside
a cone in the $\left(n',n\right)$ plane. The conjugation with $\hat{A}$
gives the multiplicative factor $e^{+a\left(\left|n\right|-\left|n'\right|\right)}$,
which decreases in the sector $\left|n'\right|>\left|n\right|$, but
increases in the sector $\left|n'\right|<\left|n\right|$. The expansivity
hypothesis insures that the cone is strictly included in the first
sector. Moreover, provided $a>0$ is small enough, the decrease of
$\left|\langle n'|\hat{M}|n\rangle\right|$ dominates the increase
of $e^{+a\left(\left|n\right|-\left|n'\right|\right)}$ in the sector
$\left|n'\right|<\left|n\right|$. At final result we obtain that
$\left|\langle n'|\hat{R}|n\rangle\right|$ decreases exponentially
fast for $\left|n\right|,\left|n'\right|\rightarrow\infty$.

\subsubsection{Exponential decrease of matrix elements}

From (\ref{eq:def_A_S1}) and (\ref{eq:def_R_S1}) it follows that
for any $n>0$:\[
\langle n'|\hat{R}|n\rangle=e^{+a\left(n-\left|n'\right|\right)}\langle n'|\hat{M}|n\rangle.\]

Instead of $n',n\in\mathbb{Z}$, we prefer to use {}``re-normalized''
indices for $n>0$, defined by \[
h=\frac{1}{n},\qquad\nu'=n'h=p+2\]
where $h,p$ were already defined in (\ref{eq:def_h_p}).

We can write $e^{+a\left(n-\left|n'\right|\right)}=e^{\frac{1}{h}A\left(\nu'\right)}$,
with $A\left(\nu'\right)\defi a\left(1-\left|\nu'\right|\right)$,
and Equation (\ref{eq:majoration_I}) gives the upper bound:\[
\left|\langle n'|\hat{M}|n\rangle\right|<e^{\frac{1}{h}B\left(\nu'\right)},\qquad B\left(\nu'\right)\defi\min\left(0,2\pi C\left(\nu'-b_{\textrm{min}}\right),2\pi C\left(b_{\textrm{max}}-\nu'\right)\right)\]
with\[
b_{\textrm{min}}=2+f'_{\textrm{min}}-\varepsilon,\qquad b_{\textrm{max}}=2+f'_{\textrm{max}}+\varepsilon.\]

\begin{figure}[htbp]
\begin{centering}\input{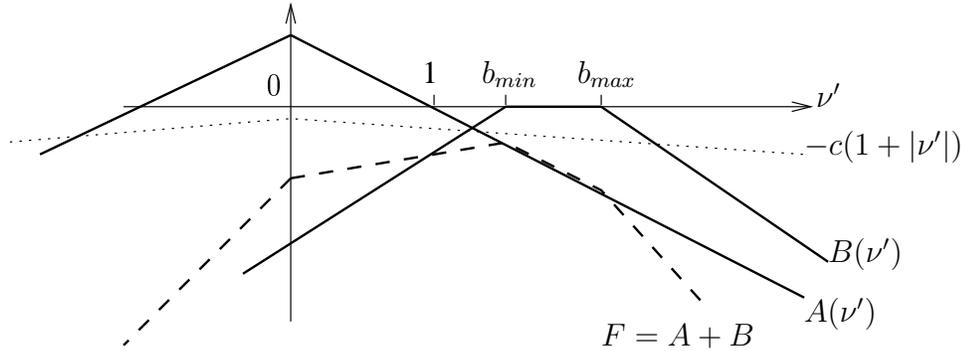}\par\end{centering}

\caption{\label{fig:A-B-F}Representation of functions $F\left(\nu'\right)=A\left(\nu'\right)+B\left(\nu'\right)$
for the upper bounds $\left|\langle n'|\hat{R}|n\rangle\right|<\exp\left(\frac{1}{h}F\left(\nu'\right)\right)<\exp\left(-\frac{c}{h}\left(1+\left|\nu'\right|\right)\right)$. }
\end{figure}

Because $M$ is expanding, we can choose $\varepsilon>0$ such that
$\min_{x}\left(M'_{x}\right)=2+f'_{\textrm{min}}>1+\varepsilon$,
hence $b_{m}>1$. For large $\left|\nu'\right|$, the functions $A\left(\nu'\right)$
and $B\left(\nu'\right)$ and have respective slope $a$ and $2\pi C$.
Choosing $a<2\pi C$ implies that the maximum of $F\left(\nu'\right)=A\left(\nu'\right)+B\left(\nu'\right)$
is reached for $\nu'=b_{m}$. See Figure \ref{fig:A-B-F}.

We need an upper bound $\left|\langle n'|\hat{R}|n\rangle\right|<\exp\left(-c\left(\left|n\right|+\left|n'\right|\right)\right)=\exp\left(-\frac{c}{h}\left(1+\left|\nu'\right|\right)\right)$,
therefore we look now for a constant $c>0$ such that $F\left(\nu'\right)=A\left(\nu'\right)+B\left(\nu'\right)<-c\left(1+\left|\nu'\right|\right)$,
for any $\nu'\in\mathbb{R}$. This requires $c<a$ and $F\left(b_{\textrm{min}}\right)\leq-c\left(1+b_{\textrm{min}}\right)\Leftrightarrow c<a\frac{b_{\textrm{min}}-1}{b_{\textrm{min}}+1}<a$.
Consequently we choose $c<a\frac{b_{\textrm{min}}-1}{b_{\textrm{min}}+1}$,
and this proves the exponential estimates of Theorem \ref{thm:R_S1}.

\subsubsection{Trace class operator}

First from $\left|\langle n'|\hat{R}|n\rangle\right|<\exp\left(-c\left(\left|n\right|+\left|n'\right|\right)\right)$,
$\hat{R}$ is a Hilbert-Schmidt operator, therefore bounded. Its domain
$C_{A}$ is dense in $L^{2}\left(S^{1}\right)$. From a classical
result, $\hat{R}$ extend in a unique way to a bounded operator in
$L^{2}\left(S^{1}\right)$. Now let $\hat{B}$ be the operator diagonal
in the Fourier basis, defined by $\hat{B}|n\rangle=\frac{1}{\left|n\right|^{\alpha}}|n\rangle$,
with $\alpha>1/2$. $\hat{B}$ is a Hilbert-Schmidt operator and $\hat{C}\defi\hat{R}\hat{B}^{-1}$
is also an Hilbert-Schmidt operator, so $\hat{R}=\hat{C}\hat{B}$
being a product of two Hilbert Schmidt operators is a trace class
operator (cf. \cite{gohberg-00}, Lemma 7.2, p.67).

\subsection{\label{sub:Exponential-accumulation-of}Exponential accumulation
of R.P. resonances near zero}

In this section, we prove Theorem \ref{th:exp_accumulation}.

Let $\hat{R}$ be the trace class operator obtained in Theorem \ref{thm:R_S1},
with the estimation $\left|\langle n'|\hat{R}|n\rangle\right|<\exp\left(-c\left(\left|n\right|+\left|n'\right|\right)\right)$
on its matrix elements, with $c>0$. We first deduce an estimation
on the singular values of $\hat{R}$:

\begin{lem}
Let $\mu_{j}$, $j=0,1,\ldots$ be the non zero singular values of
$\hat{R}$ (namely the eigenvalues of the self-adjoint operator $\sqrt{\hat{R}^{\ast}\hat{R}}$),
such that $\mu_{j+1}\leq\mu_{j}$, repeated as many times as the value
of their multiplicity. Then\begin{equation}
\mu_{j}\leq C_{1}e^{-c\, j/2}\label{eq:inequality_mj}\end{equation}
with $C_{1}=\frac{2\left(1+e^{-c}\right)}{\left(1-e^{-c}\right)^{2}}$.
\end{lem}
\begin{proof}
We borrow an argument from \cite{zworski_lin_guillope_02} (in the
proof of Prop. 3.2). From the min-max theorem, \[
\mu_{j}=\min_{V\subset L^{2}\left(S^{1}\right),\,\textrm{codim}V=j}\quad\max_{v\in V,\,\left\Vert v\right\Vert =1}\left\Vert \hat{R}v\right\Vert .\]
Consider the Fourier basis $|n\rangle$,$n\in\mathbb{Z}$ and $V_{l}=\textrm{span}\left(|n\rangle\right)_{\left|n\right|>l}$,
hence $\textrm{codim}V_{l}=2l+1$. If $|v\rangle\in V_{l}$, we compute\begin{eqnarray*}
\left\Vert \hat{R}\left|v\rangle\right|\right\Vert  & = & \left\Vert \sum_{n'\in\mathbb{Z},\left|n\right|>l}|n'\rangle\langle n'|\hat{R}|n\rangle\langle n|v\rangle\right\Vert \leq\sum_{n'\in\mathbb{Z},\left|n\right|>l}\left|\langle n'|\hat{R}|n\rangle\langle n|v\rangle\right|\\
 & \leq & \left\Vert v\right\Vert \sum_{n'\in\mathbb{Z},\left|n\right|>l}\exp\left(-c\left(\left|n\right|+\left|n'\right|\right)\right)=\left\Vert v\right\Vert \left(\sum_{\left|n\right|>l}e^{-c\left|n\right|}\right)\sum_{n'\in\mathbb{Z},}e^{-c\left|n'\right|}\\
 & = & \left\Vert v\right\Vert 2S_{l+1}\left(1+2S_{1}\right)=\left\Vert v\right\Vert e^{-c\left(l+1\right)}2S_{0}\left(1+2S_{1}\right)\end{eqnarray*}
with $S_{j}\defi=\sum_{n\geq j}e^{-cn}=\frac{e^{-cj}}{1-e^{-c}}=e^{-cj}S_{0}$.
We deduce that $\mu_{2l+1}\leq e^{-c\left(l+1\right)}2S_{0}\left(1+2S_{1}\right)$,
hence for $j$ odd,$\mu_{j}\leq e^{-cj/2}e^{-c/2}2S_{0}\left(1+2S_{1}\right)<C_{1}e^{-cj/2}$
with $C_{1}=2S_{0}\left(1+2S_{1}\right)=2\left(1+e^{-c}\right)/\left(1-e^{-c}\right)^{2}$.
For $j$ even, $\mu_{j}\leq\mu_{j-1}\leq e^{-cj/2}2S_{0}\left(1+2S_{1}\right)=C_{1}e^{-cj/2}$. 
\end{proof}
There is a fundamental relation between eigenvalues $\left(\lambda_{j}\right)_{j}$
of $\hat{R}$ (sorted such that $\left|\lambda_{j+1}\right|\leq\left|\lambda_{j}\right|$,
and repeated as many times as the value of their multiplicity) and
singular values $\left(\mu_{j}\right)_{j}$ (cf. \cite{gohberg-00},
Th. 3.1 p.52):

\begin{equation}
\prod_{j=0}^{n}\left|\lambda_{j}\right|\leq\prod_{j=0}^{n}\mu_{j},\qquad n\geq0\label{eq:inequality_sv}\end{equation}

For non zero eigenvalues, define\[
l_{j}=\log\left|\lambda_{j}\right|,\qquad m_{j}=\log\mu_{j}\]
(These sequences tend to $-\infty$ as $j\rightarrow\infty$). Then
the above inequality reads $\sum_{j=0}^{n}l_{j}\leq\sum_{j=0}^{n}m_{j}$.
Eq. (\ref{eq:inequality_mj}) gives $m_{j}\leq\log C_{1}-cj/2$. We
deduce that $\sum_{j=0}^{n}l_{j}\leq\left(n+1\right)\log C_{1}-\frac{c}{2}\frac{n\left(n+1\right)}{2}$,
hence \[
\frac{1}{\left(n+1\right)}\sum_{j=0}^{n}l_{j}\leq\log C_{1}-\frac{c}{4}n\]
But $l_{n}\leq l_{j}$ for $j\leq n$, so $l_{n}\leq\frac{1}{\left(n+1\right)}\sum_{j=0}^{n}l_{j}\leq\log C_{1}-\frac{c}{4}n$,
which proves Theorem \ref{th:exp_accumulation}.

\subsection{Numerical illustrations: Sinai-Ruelle-Bowen Measure and Ruelle-Pollicott
resonances}

In order to illustrate the previous result, we discuss here some well-known
aspects of the S.R.B. measure and Ruelle-Pollicott resonances of Example
(\ref{eq:example_f}), obtained by numerical diagonalization of operator
$\hat{R}$ (in the Fourier basis).

\subsubsection{The Sinai-Ruelle-Bowen measure}

The zero Fourier mode (constant function) $|v_{0}\rangle=|n=0\rangle$
is an eigenvector of $\hat{M}$ (and thus $\hat{R}$) with eigenvalue
$\lambda_{0}=1$. It is known that expanding maps such as eq.(\ref{eq:M_on_S1})
are mixing \cite{katok_hasselblatt}, which implies as we will see
that $\lambda_{0}=1$ is an isolated eigenvalue of multiplicity $1$,
and all other eigenvalues of $\hat{R}$ are $\left|\lambda_{i}\right|<1$,
$i=1,2\ldots$. Let $|w_{0}\rangle\in L^{2}\left(S^{1}\right)$ be
the dual eigenvector, i.e., $\langle v_{0}|w_{0}\rangle=1$ and $\hat{R}^{\ast}|w_{0}\rangle=|w_{0}\rangle\Leftrightarrow\langle w_{0}|\hat{R}=\langle w_{0}|$.
So in operator norm,\[
\hat{R}^{t}\equiv|v_{0}\rangle\langle w_{0}|+\mathcal{O}\left(\left|\lambda_{1}\right|^{t}\right),\qquad\left|\lambda_{1}\right|<1.\]

\begin{figure}[htbp]
\begin{centering}\input{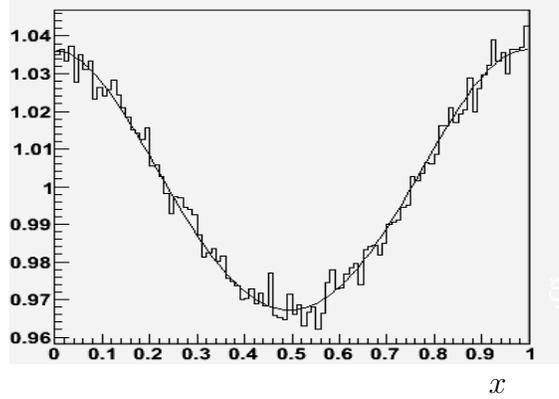}\par\end{centering}

\caption{\label{fig:mesure_SRB}S.R.B. measure: the solid line is the density
$\mu_{SRB}\left(x\right)$ computed from a numerical diagonalization
of $\hat{R}$ (in the Fourier basis), for the perturbation given in
(\ref{eq:example_f}), with $\delta=0.4$. The histogram is constructed
from a trajectory of length $T=10^{7}$ starting from a random initial
point $x_{0}$.}
\end{figure}

If $|\varphi\rangle\in L^{2}\left(S^{1}\right)$, and $|\phi\rangle\in C_{A}$,
eq.(\ref{eq:Corr_S1}) gives exponential decay of correlation for
large $t$:\begin{eqnarray*}
C_{\phi,\varphi}\left(t\right)=\langle\phi|\hat{M}^{t}|\varphi\rangle & = & \langle\phi|\hat{A}^{-1}|v_{0}\rangle\langle w_{0}|\hat{A}|\varphi\rangle+\mathcal{O}\left(\left|\lambda_{1}\right|^{t}\right),\qquad\left|\lambda_{1}\right|<1\\
 & = & \langle\phi|v_{0}\rangle\langle\mu_{SRB}|\varphi\rangle+\mathcal{O}\left(\left|\lambda_{1}\right|^{t}\right)\end{eqnarray*}
where $|\mu_{SRB}\rangle\defi\hat{A}|w_{0}\rangle\in C_{A}$ is called
the S.R.B. measure. Its density is a regular (real analytic function)
on $S^{1}$, cf Figure \ref{fig:mesure_SRB}. The physical meaning
of the last equation is that for large $t$ the function $\hat{M}^{t}|\varphi\rangle$
behaves (as seen from test functions, i.e., from a macroscopic point
of view) like the constant function $|v_{0}\rangle$ times $\langle\mu_{SRB}|\varphi\rangle$.
This is mixing property. An other interpretation of $\mu_{SRB}$ is
that for almost all $x_{0}\in S^{1}$,\[
\langle\mu_{SRB}|\equiv\lim_{T\rightarrow\infty}\frac{1}{T}\sum_{t=1}^{T}\langle\delta_{M^{t}x_{0}}|\]
(equality of measures) where $\delta_{x}$ is the Dirac measure at
$x\in S^{1}$. The right hand side is called the {}``physical measure'',
because it is constructed from a typical trajectory (\cite{eckmann_ruelle_85}
p.640, \cite{baladi_livre_00} p.73), cf Figure \ref{fig:mesure_SRB}.

\subsubsection{The Ruelle-Pollicott resonances}

Suppose for simplicity that the first $N$ eigenvalues of $\hat{R}$
are \emph{simple}, $\hat{R}|v_{i}\rangle=\lambda_{i}|v_{i}\rangle$,
$i=0\rightarrow\left(N-1\right)$, with $\lambda_{0}=1$, $\left|\lambda_{i}\right|<1$,
$\left|\lambda_{i+1}\right|\leq\left|\lambda_{i}\right|$, and $\left|\lambda_{N}\right|<\left|\lambda_{N-1}\right|$.
Let us write $|w_{i}\rangle$ the dual vectors ,i.e., $\langle w_{i}|v_{j}\rangle=\delta_{i,j}$
and $\hat{R}^{\ast}|w_{i}\rangle=\overline{\lambda}_{i}|w_{i}\rangle\Leftrightarrow\langle w_{i}|\hat{R}=\lambda_{i}\langle w_{i}|$.
Then\[
\hat{R}^{t}\equiv|v_{0}\rangle\langle w_{0}|+\sum_{i=1}^{N-1}\lambda_{i}^{t}|v_{i}\rangle\langle w_{i}|+\mathcal{O}\left(\left|\lambda_{N}\right|^{t}\right),\qquad\left|\lambda_{N}\right|<1\]
shows that the Ruelle-Pollicott resonances $\lambda_{i}$ govern the
asymptotic behaviour of the correlation functions (\ref{eq:Corr_S1})
and the convergence towards equilibrium: \[
C_{\phi,\varphi}\left(t\right)=\langle\phi|\hat{M}^{t}|\varphi\rangle=\langle\phi|v_{0}\rangle\langle\mu_{SRB}|\varphi\rangle+\sum_{i=1}^{N-1}\lambda_{i}^{t}\langle\phi|\hat{A}^{-1}|v_{i}\rangle\langle w_{i}|\hat{A}|\varphi\rangle+\mathcal{O}\left(\left|\lambda_{N}\right|^{t}\right)\]
Note that in this last expression $|v_{i}\rangle,|w_{i}\rangle\in L^{2}\left(S^{1}\right)$,
hence $\hat{A}|w_{i}\rangle\in C_{A}$ is a regular function, but
$\hat{A}^{-1}|v_{i}\rangle$ may not belong to $L^{2}\left(S^{1}\right)$.
We have to interpret $\hat{A}^{-1}|v_{i}\rangle$ as a linear form
on the space $C_{A}$. Vectors $|v_{i}\rangle,|w_{i}\rangle$ depend
on the choice of operator $\hat{A}$, but eigenvalues $\lambda_{i}$,
and distributions $\hat{A}|w_{i}\rangle,\hat{A}^{-1}|v_{i}\rangle$
do not.

\begin{figure}[htbp]
\begin{centering}\input{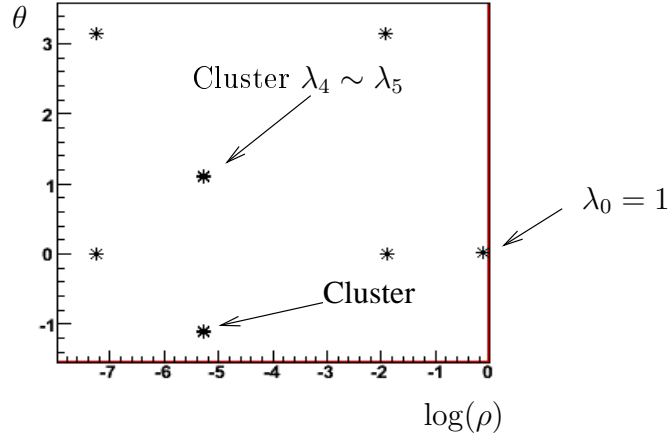}\par\end{centering}

\caption{\label{fig:RP_resonances}The first nine Ruelle-Pollicott resonances
$\lambda_{i}=\rho_{i}e^{i\theta_{i}}$ in log scale, in example Eq.(\ref{eq:example_f}),
with $\delta=0.4$. We remark two clusters of nearby eigenvalues.}
\end{figure}

Let us write $\lambda_{i}=\rho_{i}e^{i\theta_{i}},\quad\rho_{i}>0$,
hence $\log\lambda_{i}=\log\left(\rho_{i}\right)+\textrm{i}\theta_{i}$.
Figure (\ref{fig:RP_resonances}) shows the $9$ first Ruelle-Pollicott
resonances $\log\left(\lambda_{i}\right)$, obtained by a numerical
diagonalization of $\hat{R}$ in the Fourier basis. In particular,
we remark that there are two symmetric clusters of eigenvalues. $\lambda_{4}$
and $\lambda_{5}$ are very close from each other: $\log\rho_{4}=-5.271$,
$\theta_{4}=1.101$, and $\log\rho_{5}=-5.285$, $\theta_{4}=1.119$.
We have no explanation for this. In semi-classical analysis such clusters
occur because of the {}``tunnelling effect{}``. It would be nice
to find such a semi-classical interpretation here. Note that (\ref{th:exp_accumulation})
predicts that the values $\log\left(\rho_{i}\right)$ tend at least
linearly to $-\infty$, as $i\rightarrow\infty$.

\section{Hyperbolic map on the torus}

We follow essentially the same lines as in the case of an expanding
map on the circle, in order to prove Theorem \ref{thm:M_nucleaire_T2}.

\subsection{Matrix elements of the operator $\hat{M}$}

Consider the operator $\hat{M}$ defined in Eq.(\ref{eq:M_hat_T2}).
Let $|n\rangle$ denote the Fourier mode on $\mathbb{T}^{2}=\mathbb{R}^{2}/\mathbb{Z}^{2}$,
defined by $\varphi_{n}\left(x\right)=\exp\left(i2\pi n\cdot x\right)$,
with $n\in\mathbb{Z}^{2}$, $x\in\mathbb{T}^{2}$. The set $\left(|n\rangle\right)_{n\in\mathbb{Z}^{2}}$
forms a orthonormal basis of $L^{2}\left(\mathbb{T}^{2}\right)$,
and matrix elements of $\hat{M}$ in this basis are explicitly given
by:\begin{eqnarray*}
\langle n'|\hat{M}|n\rangle & = & \int_{\mathbb{T}^{2}}dx\,\exp\left(-i2\pi n'\cdot x\right)\exp\left(i2\pi n\cdot\left(M_{0}\left(x\right)+f\left(x\right)\right)\right)\\
 & = & \int_{\mathbb{T}^{2}}dx\,\exp\left(i2\pi\left(\left(M_{0}^{\mathfrak{t}}\left(n\right)-n'\right)\cdot x+n\cdot f\left(x\right)\right)\right)\end{eqnarray*}
with transposed matrix $M_{0}^{\mathfrak{t}}$.

\subsubsection{Remarks}

\begin{itemize}
\item For a vanishing perturbation $f=0$, then \begin{equation}
\langle n'|\hat{M}_{0}|n\rangle=\delta_{n'=M_{0}^{\textrm{t}}n}\label{eq:elements_matrice_M0_T2}\end{equation}
 For a non vanishing perturbation $f$, we will show now that in the
plane $n'=\left(n'_{1},n'_{2}\right)$ matrix elements are very small
outside some domain surrounding the point $n'=M_{0}^{\mathfrak{t}}n$.
\item Since $f$ is real, we have the symmetry\[
\langle-n'|\hat{M}|-n\rangle=\overline{\langle n'|\hat{M}|n\rangle}\]
and if $n=0$, we have\[
\langle n'|\hat{M}|0\rangle=\delta_{n'=0}.\]

\item With the example given by Eq.(\ref{eq:example_f_T2}), one can compute
explicitly the matrix elements in terms of the Bessel functions of
the first kind (cf \cite{abramowitz} 9.1.21 p.360):\[
\langle n'|\hat{M}|n\rangle=\left(-1\right)^{N_{1}}\delta_{N_{2}=0}\, J_{N_{1}}\left(\delta n_{2}\right),\textrm{ with}\, N=\left(n-\left(M_{0}^{-1}\right)^{\mathfrak{t}}\left(n'\right)\right)=\left\{ \begin{array}{c}
N_{1}=n_{1}-n_{1}'+n_{2}'\\
N_{2}=n_{2}+n_{1}'-2n_{2}'\end{array}\right.\]

\end{itemize}

\subsubsection{Localization property of the matrix elements}

Let us make the following change of variables $\left(n,n'\right)\Leftrightarrow\left(h,\nu,p\right)$,
for $n\neq0$, \begin{equation}
\nu=\frac{n}{\left|n\right|}\in S^{1},\qquad h=\frac{1}{\left|n\right|}>0,\qquad p=h\left(n'-M_{0}^{\mathfrak{t}}\left(n\right)\right)\in\mathbb{R}^{2},\label{eq:def_h_p_nu}\end{equation}
with $\left|n\right|=\sqrt{n_{1}^{2}+n_{2}^{2}}$ and $S^{1}$ the
unit circle in Fourier space $\mathbb{R}^{2}$. Define $\hbar=h/\left(2\pi\right)$.
Any matrix element can be written as the oscillating integral\[
I_{\hbar,\nu}\left(p\right)\defi\langle n'|\hat{M}|n\rangle=\int_{\mathbb{T}^{2}}dx\,\exp\left(i\left(\nu\cdot f\left(x\right)-p\cdot x\right)/\hbar\right).\]

\begin{figure}[htbp]
\begin{centering}\input{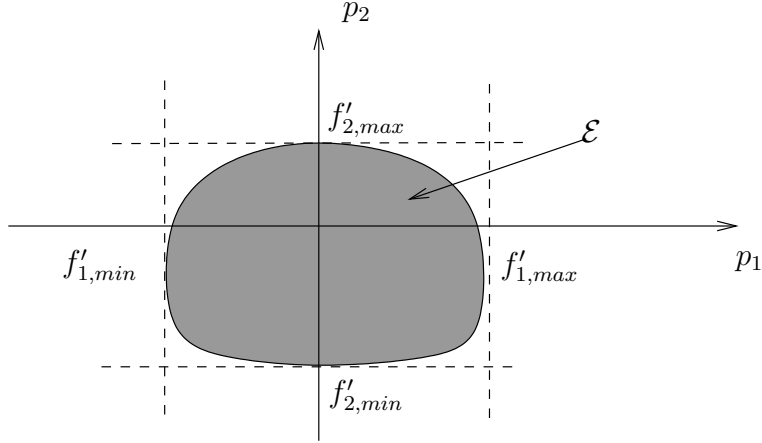}\par\end{centering}

\caption{\label{fig:domain_E}Picture of the domain $\mathcal{E}$}
\end{figure}

\begin{thm}
{}``Non stationary phase''. Let $f:\mathbb{T}^{2}\rightarrow\mathbb{R}^{2}$
be real analytic, $\hbar>0$, $\nu\in U\left(1\right)$, and $p/\left(2\pi\hbar\right)\in\mathbb{Z}^{2}$.
Then $\nu\cdot\left(D_{x}f\right)\in\mathbb{R}^{2}$. Consider the
compact domain $\mathcal{E}\defi\left\{ \nu\cdot\left(D_{x}f\right)\textrm{ s.t. }x\in\mathbb{T}^{2},\nu\in S^{1}\right\} \subset\mathbb{R}^{2}$
which contains $0$. We denote by $\left[f'_{1,\textrm{min}},f'_{1,\textrm{max}}\right]$
and $\left[f'_{2,\textrm{min}},f'_{2,\textrm{max}}\right]$ the projections
of $\mathcal{E}$ the axis $p_{1}$ and $p_{2}$ respectively . See
Figure \ref{fig:domain_E}. For any $\varepsilon>0$, there exists
$C>0$, such that for any $p=\left(p_{1},p_{2}\right)$ with $p_{1}<f'_{1,\textrm{min}}-\varepsilon$,
any $\hbar>0$ and any $\nu\in S^{1}$, one has \begin{equation}
\left|I_{\hbar,\nu}\left(p\right)\right|\leq e^{-C\left(f'_{1,\textrm{min}}-p_{1}-\varepsilon\right)/\hbar}\label{eq:majoration_I_T2}\end{equation}
Similarly we have exponential upper bounds for the other three half-planes
$p_{1}>f'_{1,\textrm{max}}+\varepsilon$, $p_{2}<f'_{2,\textrm{min}}-\varepsilon$
and $p_{2}>f'_{2,\textrm{max}}+\varepsilon$ with the same constant
$C$. Moreover we have always the general bound $\left|I_{\hbar,\nu}\left(p\right)\right|<1$.
\end{thm}

\begin{proof}
Let $\varepsilon>0$ and write $I_{\hbar,\nu}\left(p\right)=\int_{0}^{1}dx_{2}\, e^{-ip_{2}x_{2}/\hbar}\mathcal{I}_{1}\left(x_{2}\right)$
with $\mathcal{I}_{1}\left(x_{2}\right)=\int_{0}^{1}dx_{1}\, e^{i\left(\nu\cdot f\left(x_{1},x_{2}\right)-p_{1}x_{1}\right)/\hbar}$.
For $\nu,x_{2}$ fixed, let $\tilde{f}\left(x_{1}\right)=\nu\cdot f\left(x_{1},x_{2}\right)$.
Then it follows from (\ref{eq:majoration_I}) that,\[
\left|\mathcal{I}_{1}\left(x_{2}\right)\right|<e^{-C_{1}\left(\tilde{f}'_{\textrm{min}}-p_{1}-\varepsilon\right)/\hbar}\]
where $C_{1}$ and $\tilde{f}'_{\textrm{min}}$ depend on $\nu$ and
$x_{2}$. Let $f'_{1,\textrm{min}}=\textrm{min}_{\nu,x_{2}}\left(\nu.\frac{\partial f\left(x_{1},x_{2}\right)}{\partial x_{1}}\right)$
and $C=\textrm{min}_{\nu,x_{2}}\left(C_{1}\right)>0$. Then, for $p_{1}<\left(f'_{1,\textrm{min}}-\varepsilon\right)$,
one has $\left|I_{\hbar,\nu}\left(p\right)\right|<\left|\mathcal{I}_{1}\left(x_{2}\right)\right|<e^{-C\left(f'_{1,\textrm{min}}-p_{1}-\varepsilon\right)/\hbar}$.
Similarly, we define $f'_{1,\textrm{max}},f'_{2,\textrm{min}}$ and
$f'_{2,\textrm{max}}$ and obtain similar estimates.
\end{proof}

\paragraph{Remark}

\begin{itemize}
\item We have shown that the integral $I_{\hbar,\nu}\left(p\right)$ is
{}``small'' outside a rectangle containing $p=0$ in the plane $\mathbb{R}^{2}$.
This rectangle shrinks to $0$ when the perturbation $f$ is $C_{1}$
small. Coming back to variables $\left(n',n\right)$, this means that
for $n$ fixed, the matrix elements of $\langle n'|\hat{M}|n\rangle$
are {}``small'' except in a domain surrounding the point $n'=M_{0}^{\mathfrak{t}}n$. 
\end{itemize}

\subsection{\label{sub:Proof-of-Theorem_T2}Proof of Theorem \ref{thm:M_nucleaire_T2}}

Instead of the variables $n,n'\in\mathbb{Z}^{2}$, we prefer to use
for $n\neq0$,\[
h=\frac{1}{\left|n\right|}>0,\quad\nu=\frac{n}{\left|n\right|}\in S^{1},\quad\nu'=\frac{n'}{\left|n\right|}\in\mathbb{R}^{2}.\]
From (\ref{eq:def_h_p_nu}), we have $p=\nu'-M_{0}^{\textrm{t}}\nu$.
From (\ref{eq:def_A_T2}) and (\ref{eq:def_R_T2}), we have for $n\neq0$:\[
R_{n',n}\defi\langle\hat{A}n'|\hat{M}|\hat{A}^{-1}n\rangle=\exp\left(\frac{1}{h}A\left(\nu'\right)\right)\langle n'|\hat{M}|n\rangle\]
with\[
A\left(\nu'\right)=a\left(\left|\nu_{u}\right|-\left|\nu_{s}\right|-\left|\nu'_{u}\right|+\left|\nu'_{s}\right|\right),\]
where $n,h,\nu$ are considered as fixed in the discussion. Note that
we do not use yet the notation $\langle n'|\hat{R}|n\rangle$, but
$R_{n',n}$ instead, because we don't know yet if $|n\rangle$ belongs
to the domain of $\hat{R}$. This will be proved a few lines below.

\begin{figure}[htbp]
\begin{centering}\input{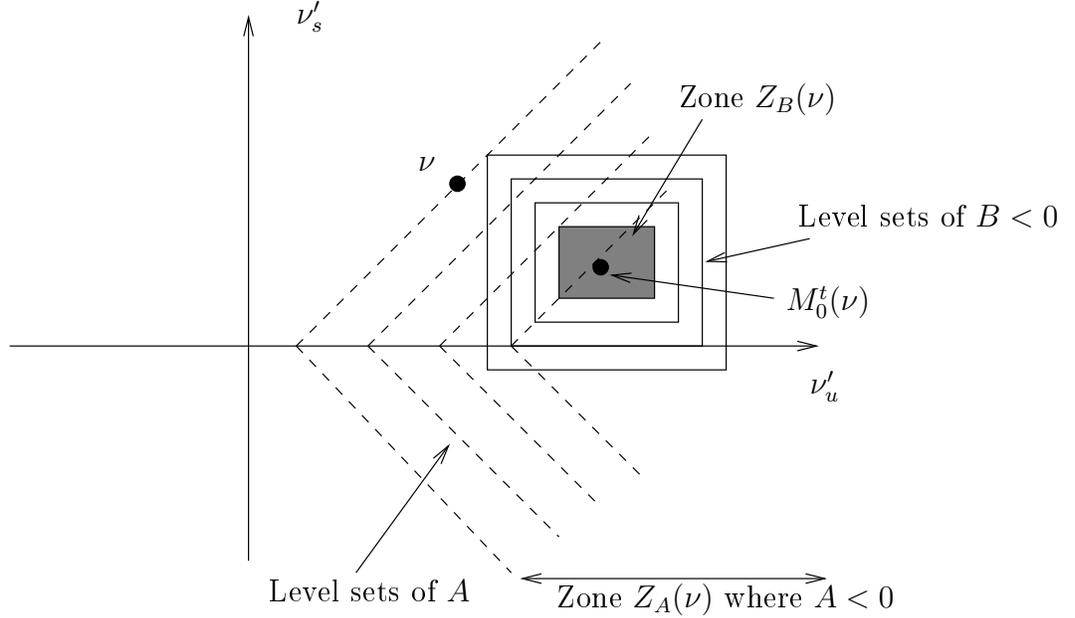}\par\end{centering}

\caption{\label{fig:ZA_ZB}Schematic representation of level sets of functions
$A\left(\nu'\right)$ and $B\left(\nu'\right)$, with respect to the
frame of unstable/stable directions. Here $\left|\nu\right|=1$, and
$\nu$ is sent to $M_{0}^{\textrm{t}}\left(\nu\right)$ by the dynamics.}
\end{figure}

Then Equation (\ref{eq:majoration_I_T2}) gives the upper bound: $\left|\langle n'|\hat{M}|n\rangle\right|<\exp\left(\frac{1}{h}B\left(\nu'\right)\right)$,
where the function $B\left(\nu'\right)$ is equal to $0$ on a rectangle
domain denoted by $Z_{B}\left(\nu\right)$, containing the point $M_{0}^{\textrm{t}}\nu$.
The function $B\left(\nu'\right)$ decreases linearly outside this
rectangle, with a slope $2\pi C$ which does not depend on $\nu$
and $h$, see Figure \ref{fig:ZA_ZB}. The size of the domain $Z_{B}\left(\nu\right)$
goes to $0$, whenever $\left\Vert f\right\Vert _{C^{1}}\rightarrow0$. 

We deduce that $\left|R_{n',n}\right|<\exp\left(\frac{1}{h}F\left(\nu'\right)\right)$,
with $F\left(\nu'\right)=A\left(\nu'\right)+B\left(\nu'\right)$.
The function $A\left(\nu'\right)$ is zero for $\nu'=\nu$, and it
is negative and decreases with a constant slope $a$ on a domain denoted
by $Z_{A}\left(\nu\right)$, see Figure \ref{fig:ZA_ZB}. At the point
$\nu'=M_{0}^{\mathfrak{t}}\left(\nu\right)=\left(e^{\lambda_{0}}\nu_{u},e^{-\lambda_{0}}\nu_{s}\right)$,
the value of $A\left(\nu'\right)=-a\left(\left|\nu_{u}\right|\left(e^{\lambda_{0}}-1\right)+\left|\nu_{s}\right|\left(1-e^{-\lambda_{0}}\right)\right)<\mathcal{A}<0$
is strictly negative, uniformly with respect to $\nu\in S^{1}$. Therefore,
the domain $Z_{B}\left(\nu\right)$ is strictly included in $Z_{A}\left(\nu\right)$
if the perturbation $f$ is small enough in $C^{1}$ norm.

If we choose $a$ such that $A$ increases slower than $B$ decreases
(i.e., $a<c'2\pi C$ where $c'>0$) then there exists $c>0$ such
that \[
F\left(\nu'\right)<-c\left(\left|\nu_{1}\right|+\left|\nu_{2}\right|+\left|\nu'_{1}\right|+\left|\nu'_{2}\right|\right)\]
This gives $\left|R_{n',n}\right|<\exp\left(-c\left(\left|n_{1}\right|+\left|n_{2}\right|+\left|n'_{1}\right|+\left|n'_{2}\right|\right)\right)$.

Let us first deduce that the operator $\hat{R}=\hat{A}\hat{M}\hat{A}^{-1}$
is defined on the domain $C_{A}$. If $|\phi\rangle\in C_{A}$, \begin{eqnarray*}
|\phi\rangle & \in & \textrm{Dom}\left(\hat{R}\right)\Leftrightarrow\hat{M}\hat{A}^{-1}|\phi\rangle\in D_{A}\Leftrightarrow\sum_{n'}\left|\langle\hat{A}n'|\hat{M}\hat{A}^{-1}|\phi\rangle\right|^{2}<\infty\\
 & \Leftrightarrow & \sum_{n'}\left|\sum_{n}\langle\hat{A}n'|\hat{M}|\hat{A}^{-1}n\rangle\langle n|\phi\rangle\right|^{2}<\infty\Leftrightarrow\sum_{n'}\left|\sum_{n}R_{n',n}\langle n|\phi\rangle\right|^{2}<\infty.\end{eqnarray*}

Let us show now the last estimate is actually fulfilled. If $|\phi\rangle\in L^{2}\left(\mathbb{T}^{2}\right)$
\begin{eqnarray*}
\sum_{n'}\left|\sum_{n}R_{n',n}\langle n|\phi\rangle\right|^{2} & \leq & \sum_{n'}\left(\sum_{n}\left|R_{n',n}\right|\left|\langle n|\phi\rangle\right|\right)^{2}\\
 & \leq\left\Vert \phi\right\Vert ^{2} & \sum_{n'}e^{-2c\left(\left|n'_{1}\right|+\left|n'_{2}\right|\right)}\left(\sum_{n}e^{-c\left(\left|n_{1}\right|+\left|n_{2}\right|\right)}\right)^{2}<\infty.\end{eqnarray*}
Therefore the operator $\hat{R}=\hat{A}\hat{M}\hat{A}^{-1}$ is defined
on the domain $C_{A}$ and its matrix elements are obviously $\langle n'|\hat{R}|n\rangle=R_{n',n}$.
With the same arguments as those used earlier for the expanding map
on $S^{1}$, we deduce that $\hat{R}$ extends to a trace class operator
on $L^{2}\left(\mathbb{T}^{2}\right)$.

\subsubsection{\label{sub:Exponential-accumulation-T2}Proof of the exponential
concentration of R.P. resonances}

The proof of Theorem \ref{th:exp_accumulation_T2} is very similar
to the proof we gave in Section \ref{sub:Exponential-accumulation-of}.
Here we use the same notations and emphasize the differences.

Let $\hat{R}$ be the trace class operator obtained in Theorem \ref{thm:M_nucleaire_T2},
with the estimation $\left|\langle n'|\hat{R}|n\rangle\right|<\exp\left(-c\left(\left|n_{1}\right|+\left|n_{2}\right|+\left|n'_{1}\right|+\left|n'_{2}\right|\right)\right)$
on its matrix elements, with $c>0$. 

\begin{lem}
Let $\mu_{j}$, $j=0,1,\ldots$ be the non zero singular values of
$\hat{R}$, such that $\mu_{j+1}\leq\mu_{j}$, repeated as many times
as the value of their multiplicity. Then\begin{equation}
\mu_{j}\leq C_{1}e^{-\frac{c}{2}\,\sqrt{j}}\label{eq:inequality_mj_T2}\end{equation}
with $C_{1}>0$.
\end{lem}
\begin{proof}
From the min-max theorem, \[
\mu_{j}=\min_{V\subset L^{2}\left(\mathbb{T}^{2}\right),\,\textrm{codim}V=j}\quad\max_{v\in V,\,\left\Vert v\right\Vert =1}\left\Vert \hat{R}v\right\Vert .\]
Consider the Fourier basis $|n\rangle$,$n=\left(n_{1},n_{2}\right)\in\mathbb{Z}^{2}$
and $V_{l}=\textrm{span}\left(|n\rangle\right)_{\max\left(\left|n_{1}\right|,\left|n_{2}\right|\right)>l}$,
hence $\textrm{codim}V_{l}=\left(2l+1\right)^{2}$. If $|v\rangle\in V_{l}$,
we compute\begin{eqnarray*}
\left\Vert \hat{R}\left|v\rangle\right|\right\Vert  & = & \left\Vert \sum_{n'\in\mathbb{Z},\left|n\right|>l}|n'\rangle\langle n'|\hat{R}|n\rangle\langle n|v\rangle\right\Vert \\
 & \leq & \left\Vert v\right\Vert \left(\sum_{n'\in\mathbb{Z}^{2}}\exp\left(-c\left(\left|n'_{1}\right|+\left|n'_{2}\right|\right)\right)\right)\left(\sum_{n/\max\left(\left|n_{1}\right|,\left|n_{2}\right|\right)>l}\exp\left(-c\left(\left|n_{1}\right|+\left|n_{2}\right|\right)\right)\right)\\
 & \leq & \left\Vert v\right\Vert Ce^{-cl},\qquad C>0\end{eqnarray*}
We deduce that $\mu_{j}\leq C_{1}e^{-c\sqrt{j}/2}$, with $C_{1}>0$.
\end{proof}
For non zero eigenvalues, define\[
l_{j}=\log\left|\lambda_{j}\right|,\qquad m_{j}=\log\mu_{j}\]
(These sequences tend to $-\infty$ as $j\rightarrow\infty$). Inequality
(\ref{eq:inequality_sv}) reads $\sum_{j=0}^{n}l_{j}\leq\sum_{j=0}^{n}m_{j}$.
Eq. (\ref{eq:inequality_mj_T2}) gives $m_{j}\leq\log C_{1}-c\sqrt{j}/2$.
We deduce that $\sum_{j=0}^{n}l_{j}\leq\left(n+1\right)\log C_{1}-\frac{c}{2}\sum_{j=0}^{n}\sqrt{j}$.
But $\sum_{j=0}^{n}\sqrt{j}\geq\int_{0}^{n}\sqrt{x}dx=\frac{2}{3}n^{3/2}$.
Hence $\frac{1}{\left(n+1\right)}\sum_{j=0}^{n}l_{j}\leq\log C_{1}-\frac{c}{3}\frac{n^{3/2}}{\left(n+1\right)}$.
But $l_{n}\leq l_{j}$ for $j\leq n$, so $l_{n}\leq\frac{1}{\left(n+1\right)}\sum_{j=0}^{n}l_{j}\leq\log C_{1}-\frac{c}{3}\frac{n^{1/2}}{\left(1+1/n\right)}$,
which proves Theorem \ref{th:exp_accumulation_T2}.

\section{Conclusion}

For specific models of chaotic dynamics, namely real analytic expanding
maps on the circle $S^{1}$ and real analytic hyperbolic maps on the
torus $\mathbb{T}^{2}$, we have shown that the decay of time correlation
functions can be described by a trace class operator in $L^{2}\left(S^{1}\right)$
(resp. $L^{2}\left(\mathbb{T}^{2}\right)$). We have followed an approach
similar to the {}``complex scaling method'', well-known to study
the decay of quantum states in open quantum systems. As explained
in the introduction, this approach has been already used by V. Baladi
and M Tsujii \cite{baladi_05} for hyperbolic diffeomorphisms in a
more general context, but our methods differ sightly and allowed us
to obtain different results. To make a more precise comparison, our
operator $\hat{A}$ defined in Eq.(\ref{eq:def_A_T2}) and the conjugation
in Eq.(\ref{eq:def_R_T2}), correspond to their definition of anisotropic
norms. But they use powers of the Fourier components whereas we use
exponential of them. This exponential is important for us to obtain
a trace class operator $\hat{R}$, as explained in section \ref{sub:Remarks-313}. 

In this paper, some {}``semi-classical aspects'' of hyperbolic dynamics
have appeared many times: (i) in Theorem \ref{thm:Non-stationary-phase.}
concerning the localization of the matrix elements in Fourier space,
and the remark which follows, (ii) in a remark on the semi-classical
Weyl law after equation (\ref{eq:asympt_li_T2}), and (iii) with Figure
\ref{fig:RP_resonances} where a cluster of eigenvalues suggests some
semi-classical tunnelling effect. The role of semi-classical parameter
is played by the inverse of the distance in Fourier space: $\hbar\equiv1/\left|n\right|$.
A direction of research would be to make this semi-classical theory
more precise. 

We would like to comment some limitations of our results and possible
extensions of them. First we have assumed that the dynamics is given
by a real analytic map. This is a severe limitation because in hyperbolic
dynamical system theory one has to use Hölder potential functions
\cite{baladi_livre_00}\cite{katok_hasselblatt}. We have presented
here the results in their simplest form. In particular, for expanding
maps on the torus, we have shown that $\hat{R}$ is trace class for
any expanding map, but for hyperbolic maps on the torus we have assumed
that the non linear perturbation $f$ is weak enough. In geometric
terms we have supposed that $\left\Vert f\right\Vert _{C^{1}}$ is
weak enough so that the unstable and stable foliations are respectively
contained in fixed cones (the cones adapted to the linear map, and
which enter in the definition of $\hat{A}$ eq.(\ref{eq:def_A_T2})).
It could be possible to generalize in this direction and treat in
this way \emph{any} uniform analytic hyperbolic map on the torus,
using a local choice of cones. As in \cite{baladi_05}, this could
be possible using pseudo-differential operators instead of $\hat{A}$.
Then the localization property of the matrix elements in Theorem \ref{thm:Non-stationary-phase.}
would be replaced by a {}``microlocal'' version in the cotangent
space $T^{*}S^{1}$(resp. $T^{*}\mathbb{T}^{2}$). Some other directions
of research could be to take advantage of the relative simplicity
of this approach to investigate non uniform hyperbolic dynamics or
other kinds of dynamical systems which exhibit some chaotic behaviour,
where many questions remain open.

\bibliographystyle{plain}
\bibliography{/home/faure/articles/articles}

\end{document}